\documentclass[11pt]{article}

\usepackage{amsmath}
\usepackage{amssymb}
\usepackage{amsfonts}
\usepackage{latexsym}
\usepackage{amsthm}
\usepackage{palatino}
\usepackage{mathpazo}
\usepackage{stmaryrd}
\usepackage[T1]{fontenc}
\usepackage{textcomp}
\usepackage{dsfont}
\usepackage{cite}
\usepackage{graphicx}

\newtheorem{theorem}{Theorem}

\newtheorem{corollary}{Corollary}

\theoremstyle{definition}

\newcommand*{\cX}{\mathcal{X}}
\newcommand*{\cY}{\mathcal{Y}}

\newcommand{\ket}[1]{|#1\rangle}
\newcommand{\bra}[1]{\langle #1 |}

\newcommand{\beq}{\begin{equation}}
\newcommand{\eeq}{\end{equation}}
\newcommand{\best}{\begin{equation*}}
\newcommand{\eest}{\end{equation*}}
\newcommand{\tinyspace}{\mspace{1mu}}

\newcommand{\op}[1]{\operatorname{#1}}

\newcommand{\norm}[1]{\left\lVert\tinyspace#1\tinyspace\right\rVert}
\newcommand{\snorm}[1]{\lVert\tinyspace#1\tinyspace\rVert}
\newcommand{\abs}[1]{\left\lvert\tinyspace #1 \tinyspace\right\rvert}

\newcommand{\defeq}{\stackrel{\smash{\text{\tiny def}}}{=}}
\newcommand{\tr}{\operatorname{Tr}}

\renewcommand{\t}{{\scriptscriptstyle\mathsf{T}}}
\newcommand{\ip}[2]{\left\langle #1 , #2\right\rangle}

\def\({\left(}
\def\){\right)}

\def\I{\mathds{1}}

\newcommand{\setft}[1]{\textup{#1}}
\newcommand{\lin}[1]{\setft{L}\left(#1\right)}
\newcommand{\density}[1]{\setft{D}\left(#1\right)}
\newcommand{\unitary}[1]{\setft{U}\left(#1\right)}
\newcommand{\trans}[1]{\setft{T}\left(#1\right)}
\newcommand{\herm}[1]{\setft{Herm}\left(#1\right)}
\newcommand{\pos}[1]{\setft{Pos}\left(#1\right)}

\newcommand{\sep}[1]{\setft{Sep}\left(#1\right)}
\newcommand{\sepd}[1]{\setft{SepD}\left(#1\right)}

\newcommand{\ppt}[1]{\setft{PPT}\left(#1\right)}

\def\complex{\mathbb{C}}

\newcommand{\reg}[1]{\mathsf{#1}}

\newenvironment{mylist}[1]{\begin{list}{}{
	\setlength{\leftmargin}{#1}
	\setlength{\rightmargin}{0mm}
	\setlength{\labelsep}{2mm}
	\setlength{\labelwidth}{8mm}
	\setlength{\itemsep}{0mm}}}
	{\end{list}}

\def\X{\mathcal{X}}
\def\Y{\mathcal{Y}}
\def\Z{\mathcal{Z}}

\begin{document}

\title{\textbf{Entanglement in channel discrimination\\
    with restricted measurements}}

\author{
  William Matthews$^\ast$ 
  \hspace{7mm}
  Marco Piani$^\dagger$
  \hspace{7mm}
  John Watrous$^\ddagger$\\[3mm]
 $^\ast$%
  {\small\it Institute for Quantum Computing and Department of
    Combinatorics and Optimization}\\[-1mm]
  {\small\it University of Waterloo}\\[-1mm]
  {\small\it Waterloo, Ontario, Canada}\\
  $^\dagger$%
  {\small\it Institute for Quantum Computing and Department of Physics
    and Astronomy}\\[-1mm]
  {\small\it University of Waterloo}\\[-1mm]
  {\small\it Waterloo, Ontario, Canada}\\
  $^\ddagger$%
  {\small\it Institute for Quantum Computing and School of Computer
    Science}\\[-1mm]
  {\small\it University of Waterloo}\\[-1mm]
  {\small\it Waterloo, Ontario, Canada}
}

\maketitle

\begin{abstract}
  We study the power of measurements implementable with local quantum
  operations and classical communication (or \emph{LOCC measurements}
  for short) in the setting of quantum channel discrimination.
  More precisely, we consider discrimination procedures that attempt
  to identify an unknown channel, chosen uniformly from two known
  alternatives, that take the following form:
  (i) the input to the unknown channel is prepared in a possibly
  entangled state with an ancillary system, (ii) the unknown channel
  is applied to the input system, and (iii) an LOCC measurement is
  performed on the output and ancillary systems, resulting in a guess
  for which of the two channels was given.
  The restriction of the measurement in such a procedure to be an LOCC
  measurement is of interest because it isolates the entanglement in
  the initial input/ancillary systems as a resource in the setting of
  channel discrimination.
  We prove that there exist channel discrimination problems for which
  restricted procedures of this sort can be at either of the two
  extremes: they may be optimal within the set of all discrimination
  procedures (and simultaneously outperform all strategies that make
  no use of entanglement), or they may be no better than unentangled
  strategies (and simultaneously sub-optimal within the set of all
  discrimination procedures).
\end{abstract}

%--------------------------------------------------------------------------%
\section{Introduction}
%--------------------------------------------------------------------------%

Quantum channel discrimination is an interesting problem in the theory
of quantum information.
In this problem, two known physical processes (or channels) are fixed,
and access to one of them is made available---but it is not known
which one it is.
In the simplest scenario only a single application of the channel is
possible.
The goal is to determine, with minimal probability of error, which of
the two channels was given, assuming for simplicity that the two
channels were equally likely.
Several papers, including
\cite{ChildsPR00,
  D'ArianoPP01,
  Acin01,
  DuanFY09,
  GiovannettiLM04,
  GilchristLN05,
  RosgenW05,
  Sacchi05a,
  Sacchi05b,
  Lloyd08,
  Rosgen08,
  Watrous08},
have considered variants of this problem.

The most general form of a discrimination procedure for a channel
discrimination problem of the type described above has the following
form:
(i) the input to the unknown channel is prepared in a possibly
entangled state with an ancillary system, (ii) the unknown channel
is applied to the input system, and (iii) a measurement is
performed on the output and ancillary systems, resulting in a guess
for which of the two channels was given.
The primary purpose of this paper is to consider the effect of
restricting the measurements in step (iii) to be LOCC (i.e.,
implementable using only local operations and classical communication
with respect to the output/ancillary system splitting).

It is well-known that entanglement between the input and ancillary
systems is sometimes advantageous for channel discrimination, in the
sense that it may allow for a strictly smaller probability of error
to correctly identify the given channel in comparison to the case
where there is no entanglement (which turns out to be equivalent to
having no ancillary system whatsoever).
This phenomenon seems to have been identified first by
Kitaev~\cite{Kitaev97}, who introduced the \emph{diamond norm} on
super-operators to deal with precisely this phenomenon.
More recent work, partially represented by the sources cited above,
has further illuminated the usefulness of entanglement in the problem
of channel discrimination and related tasks.

In~\cite{PianiW09} the question was reversed, by supposing that some
arbitrary entangled state is given and asking whether the
entanglement in this state is useful for channel discrimination. 
It was proved that every bipartite entangled state indeed does provide
an advantage for this task: 
there necessarily exists an instance of a channel discrimination
problem for which the entangled state allows for a correct
discrimination with strictly higher probability than every possible
unentangled (or separable) state. 
One may therefore say that every entangled state is a resource for
channel discrimination.
However, there is an obvious (potential) way that this result might be
improved, which is to prove that every entangled state remains useful
in the setting of channel discrimination even when measurements are
restricted to be LOCC as suggested above.
This would have the effect of isolating the entanglement in the
input/ancillary systems as the principal use of entanglement in such a
procedure, and therefore as a more fundamentally important resource.
We have not been able to determine whether or not such a result holds,
and we consider this to be one of the main open problems of interest
associated with this work.

Restrictions on operations are standard in entanglement
theory~\cite{HorodeckiHHH09}, where they are introduced both because
they are physically motivated (e.g., some initial entanglement between
distant labs may be established by means of optical fibers or flying
photons, but beyond this only local operations and classical
communication may be feasible) 
and because they make entanglement theory an interesting resource
theory, where the specifically quantum properties of entanglement are
emphasized.
The effect of restrictions on the measurement on the ability to
discriminate---either ambiguously or unambiguously---between quantum
states has also recently attracted much attention. 
In particular, the limits of LOCC discrimination have been
investigated in
\cite{WalgateSHV00,
  VirmaniSPM01,
  WalgateH02,
  Fan04,
  HorodeckiOSS04,
  chefles04,
  GhoshJKKR05,
  Watrous05b,
  Nathanson05,
  HayashiMT06,
  hayashiMMOV06,
  Fan07,
  DuanFJY07,
  OwariH08,
  BandyopadhyayW09,
  MatthewsW09},
for instance.
Such limits are at the base of the existence of hiding
states~\cite{TerhalDL01,EggelingW02,DivincenzoLT02,MatthewsW09},
which are orthogonal and therefore perfectly distinguishable by global
operations, but hardly distinguishable by LOCC measurements. 
A systematic approach to investigating the relation between
distinguishability of states under various restrictions on
measurements was put forward in~\cite{MatthewsWW09}. 

In regard to the role of measurements and restrictions on measurements
in quantum channel discrimination, the focus of past work has mainly
been on the discrimination of bi- and multi-partite unitary operations.
It has been proved that with many uses of the unknown unitary, it is
possible to perfectly discriminate between any set of unitaries, as
long as the parties can apply an LOCC protocol~\cite{ZhouZG07,DuanFY08}. 
We stress that the unitaries acts globally on the same parties on
which the restrictions are imposed. 
The issue we consider in this paper---the role of restrictions on the
measurements in minimum-error channel discrimination---deals with
concepts similar to those present in~\cite{ZhouZG07,DuanFY08}, but with
critical differences.
Indeed, in our case we consider that (i) only one use of the unknown
channel is allowed, and we do not focus on perfect discrimination;
and (ii) the channels---possibly unitaries---are applied only to the
input, so the input/ancilla evolution is local. 
In particular, in the case of unitaries it is known that input/ancilla
entanglement is useless for minimum-error channel discrimination, so
restrictions on the measurements that are based on input/ancilla
locality are uninteresting.

The main contribution of the present paper is the identification of
instances of channel discrimination problems where procedures
restricted to make LOCC measurements are at either of the two
extremes: they may be optimal within the set of all discrimination
procedures and simultaneously outperform all strategies that make
no use of entanglement, or they may be no better than unentangled
strategies and simultaneously sub-optimal within the set of all
discrimination procedures.
These two possibilities are discussed in Sections~\ref{sec:useful} and
\ref{sec:useless}, respectively, which follow
Section~\ref{sec:definitions} that provides some definitions that are
useful for describing the examples.
The paper concludes with Section~\ref{sec:conclusion}, which discusses
some future research directions relating to our work.

%--------------------------------------------------------------------------%
\section{Definitions}
\label{sec:definitions}
%--------------------------------------------------------------------------%

Throughout this paper we will use notation and terminology that, for
the most part, is standard in the theory of quantum information.
For the sake of clarity let us state explicitly that we restrict our
attention to finite-dimensional complex Hilbert spaces in this paper,
and for any such space $\X$ we write $\lin{\X}$, $\herm{\X}$,
$\pos{\X}$ and $\density{\X}$ to denote the sets of all linear
operators, Hermitian operators, positive semidefinite operators, and
density operators on $\X$, respectively.
We also write $\sep{\X:\Y}$ and $\ppt{\X:\Y}$ to denote the sets of
all (unnormalized) separable and PPT (i.e., positive partial
transpose) operators on a tensor product space $\X\otimes\Y$.
\footnote{%
 Both $\sep{\X:\Y}$ and $\ppt{\X:\Y}$ are subsets of $\pos{\X\otimes\Y}$. An positive operator is PPT if it remains positive under the action of partial transposition;
 a positive operator is separable if it can be expressed as $\sum_iP_i\otimes Q_i$, with $P_i\in\pos{\X},Q_i\in\pos{\Y}$.}

When we refer to a \emph{channel} we mean a completely positive,
trace-preserving linear mapping of the form
\begin{equation} \label{eq:mapping}
\Phi:\lin{\X}\rightarrow\lin{\Y}.
\end{equation}
Hereafter we will write $\trans{\X,\Y}$ to refer to the vector space
of all (not necessarily completely positive or trace-preserving)
mappings of the form \eqref{eq:mapping}.

As described in the introduction, this paper concerns the problem of
channel discrimination.
The specific type of channel discrimination problems we consider are
as follows.
Two channels $\Phi_0,\Phi_1\in\trans{\X,\Y}$ are fixed.
One of the two channels is selected, uniformly at random, and a single
evaluation of this (unknown) channel is made available.
The goal is to determine which of the two channels was selected.

A natural, but sometimes sub-optimal, strategy for solving an instance
of a channel discrimination problem is to choose a quantum state
$\rho\in\density{\X}$, to apply the unknown channel to $\rho$, and to
measure the resulting state according to a binary-valued measurement
$\{P_0,P_1\}\subset\pos{\Y}$.
The measurement outcome (0 or 1) is then interpreted as the
procedure's guess for which channel was given.
The probability that such a procedure correctly identifies the unknown
channel is given by
\[
\frac{1}{2}\ip{P_0}{\Phi_0(\rho)} + 
\frac{1}{2}\ip{P_1}{\Phi_1(\rho)}
=
\frac{1}{2} + \frac{1}{4} \ip{P_0-P_1}{\Phi_0(\rho)-\Phi_1(\rho)}
\]
where the inner product is the Hilbert-Schmidt inner product:
$\ip{X}{Y} = \tr(X^{\ast}Y)$.
Optimizing over all choices of $\rho\in\density{\X}$ and all
binary-valued measurements $\{P_0,P_1\}$ on $\Y$ yields a correctness
probability
\[
\frac{1}{2} + \frac{1}{4}\norm{\Phi_0 - \Phi_1}_{\textup{NE}},
\]
where the norm $\norm{\Phi}_{\textup{NE}}$ is defined as
\[
\norm{\Phi}_{\textrm{NE}}
= \max_{\rho\in\density{\X}} \norm{\Phi(\rho)}_1
\]
for every Hermiticity-preserving\footnote{A mapping
  $\Phi\in\trans{\X,\Y}$ is \emph{Hermiticity-preserving} if and only
  if $\Phi(X)\in\herm{\Y}$ for every $X\in\herm{\X}$.
  For any choice of quantum channels $\Phi_0,\Phi_1\in\trans{\X,\Y}$
  it holds that $\Phi = \Phi_0 - \Phi_1$ is Hermiticity-preserving.}
mapping $\Phi\in\trans{\X,\Y}$.
This norm could be extended to arbitrary (non-Hermiticity-preserving)
mappings, but it is not necessary for us to consider such extensions
in this paper.
The subscript $\textrm{NE}$ for this norm is short for ``no
entanglement,'' which refers to the fact that the discrimination
procedure has not made use of the possibility that the input system to
the unknown channel could have been entangled with an ancillary
system.

A more general type of discrimination strategy that does make use
of an ancillary system is as follows.
A quantum state $\rho\in\density{\X\otimes\Z}$ (for an arbitrary
choice of $\Z$) is selected, and the unknown channel is applied to the
part of this state corresponding to $\X$.
A binary-valued measurement $\{P_0,P_1\}\subset\pos{\Y\otimes\Z}$ is
then applied to the resulting state, and (as before) the outcome is
interpreted as the procedure's guess for which channel was given.
The probability for such a procedure to correctly identify the unknown
channel is
\begin{equation} \label{eq:inner-product-probability}
\frac{1}{2} + \frac{1}{4}
\ip{P_0-P_1}{(\Phi_0\otimes\I_{\lin{\Z}})(\rho)-
(\Phi_1\otimes\I_{\lin{\Z}})(\rho)}.
\end{equation}
Optimizing over all choices of $\rho\in\density{\X\otimes\Z}$ and all
binary-valued measurements $\{P_0,P_1\}$ on $\Y\otimes\Z$, for
any choice of $\Z$ having dimension at least that of $\X$, results in
the quantity
\[
\frac{1}{2} + \frac{1}{4}\norm{\Phi_0 - \Phi_1}_{\diamondsuit},
\]
where the \emph{diamond norm} $\norm{\cdot}_{\diamondsuit}$ is defined
as
\begin{equation} \label{eq:diamond-norm}
\norm{\Phi}_{\diamondsuit} = 
\max_{\rho\in\density{\X\otimes\X}}
\norm{(\Phi\otimes\I_{\lin{\X}})(\rho)}_1
\end{equation}
for every Hermiticity-preserving mappings $\Phi\in\trans{\X,\Y}$.
(For general maps, the maximum is taken over all
$X\in\lin{\X\otimes\X}$ with $\norm{X}_1\leq 1$.)

It is known that this more general sort of strategy can give a
striking improvement in the probability to correctly discriminate some
pairs of channels.
It is the \emph{entanglement} between the input and ancillary systems
that is responsible for this improvement, when it occurs, because (as
observed in \cite{PianiW09}) it holds that
\[
\max_{\rho\in\sepd{\X:\Z}} \norm{(\Phi\otimes \I_{\lin{\Z}})(\rho)}_1
= \norm{\Phi}_{\textrm{NE}}
\]
for every choice of $\Z$, where $\sepd{\X:\Z}$ denotes the set of
separable density operators on $\X\otimes\Z$.
In other words, the ancillary system is useless for channel
discrimination unless it is entangled with the input system.

With the connection between channel discrimination and the two norms
defined above in mind, we define a norm $\norm{\Phi}_{\textrm{LOCC}}$
by considering a maximization of the expression
\eqref{eq:inner-product-probability} over those choices of $P_0$ and
$P_1$ that represent \emph{LOCC measurements}, as opposed to general
measurements.
More precisely, for any Hermiticity-preserving mapping
$\Phi\in\trans{\X,\Y}$ we define
\begin{equation} \label{eq:LOCC-norm}
  \norm{\Phi}_{\textrm{LOCC}}
  = \max_{\{P_0,P_1\}} \max_{\rho\in\density{\X\otimes\Z}}
  \ip{P_0-P_1}{(\Phi\otimes\I_{\lin{\Z}})(\rho)},
\end{equation}
where the maximization is taken over all LOCC measurements\footnote{%
  It is not known if the set of LOCC measurements on a bipartite
  system $\Y\otimes\Z$, implementable with a finite number of rounds
  of communication, is closed.
  For the sake of simplicity, we will consider any measurement in the
  closure of the set of LOCC measurements to be an LOCC measurement,
  so that the maximum over $\{P_0,P_1\}$ in \eqref{eq:LOCC-norm} is a
  maximization over a compact set.}
$\{P_0,P_1\}$ on $\Y\otimes\Z$, and where $\Z$ is any space having
dimension at least that of $\X$.
We also define $\norm{\Phi}_{\textrm{SEP}}$ and
$\norm{\Phi}_{\textrm{PPT}}$ similarly, where the maximization is over
all separable or PPT binary-valued measurements, respectively.\footnote{%
 A binary-valued measurement $\{P_0,P_1\}$ is separable or PPT if both $P_0$ and $P_1$ are
 separable or PPT, respectively.}
In all of these cases, the resulting norm is insensitive to the
dimension of $\Z$, provided it is at least that of $\X$.
This follows from the observation that the maximum over
$\rho\in\density{\X\otimes\Z}$ for any of these norms is always
achieved for a pure state (by a simple convexity argument),
and such a state must be supported on a subspace of $\Z$ having
dimension at most that of $\X$.
Similar to $\norm{\Phi}_{\textup{NE}}$, we do not concern ourselves
with extensions of $\norm{\Phi}_{\textrm{LOCC}}$, 
$\norm{\Phi}_{\textrm{SEP}}$, or $\norm{\Phi}_{\textrm{PPT}}$ to
non-Hermiticity-preserving mappings $\Phi$.

It is clear that
\begin{equation} \label{eq:norm-relations}
\norm{\Phi}_{\textrm{NE}} \leq
\norm{\Phi}_{\textrm{LOCC}} \leq
\norm{\Phi}_{\textrm{SEP}} \leq
\norm{\Phi}_{\textrm{PPT}} \leq
\norm{\Phi}_{\diamondsuit}
\end{equation}
and the main contribution of this paper is to provide examples of
channels $\Phi_0$ and $\Phi_1$ for which the mapping 
$\Phi = \Phi_0 - \Phi_1$ causes various choices of the inequalities in
\eqref{eq:norm-relations} to become either equalities or strict
inequalities.
We are, in particular, interested in the relationship between
$\norm{\Phi_0 - \Phi_1}_{\textrm{LOCC}}$ and the two norms
$\norm{\Phi_0 - \Phi_1}_{\textrm{NE}}$ and 
$\norm{\Phi_0 - \Phi_1}_{\diamondsuit}$ for different choices of
channels $\Phi_0$ and $\Phi_1$.
This relationship addresses the question raised in the introduction,
which is whether entanglement between the input and auxiliary systems
remains a useful resource for channel discrimination when entangled
measurements are disallowed. The two sections that follow show that sometimes entanglement is still
useful in this sense and sometimes it is not.

\begin{figure}[htbp]
   \centering
   \includegraphics[width=0.5\textwidth]{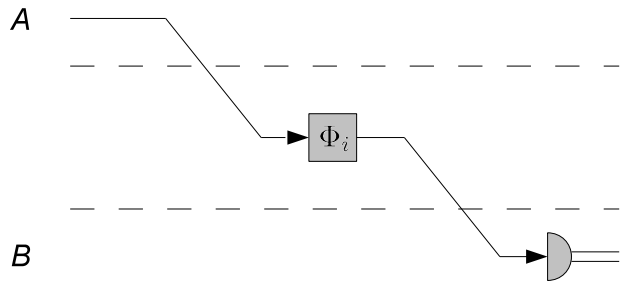} % requires the graphicx package
   \caption{Operational scenario for the NE norm. The unknown channel $\Phi_i$ goes from Alice to Bob, and it is not in the hands of either of them. Bob performs the final measurement. In this case, Alice's input to the channel is uncorrelated with any other system in the hands of Alice or Bob.}
   \label{fig:trace}
\end{figure}

\begin{figure}[htbp]
   \centering
   \includegraphics[width=0.5\textwidth]{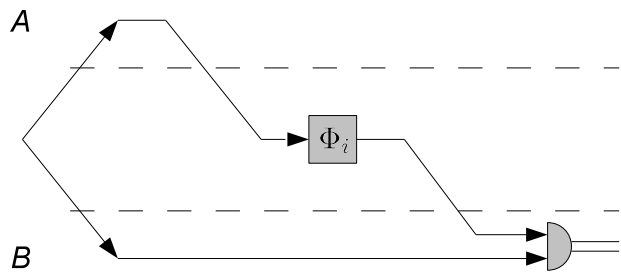} % requires the graphicx package
   \caption{Operational scenario for the diamond norm. Alice and Bob share an initial---possibly entangled---probe/ancilla state which is optimal for channel discrimination. Alice sends the probe downs the channel and Bobs proceeds to the measurement of the output probe/ancilla state.}
   \label{fig:diamond}
\end{figure}

\begin{figure}[htbp]
   \centering
   \includegraphics[width=0.5\textwidth]{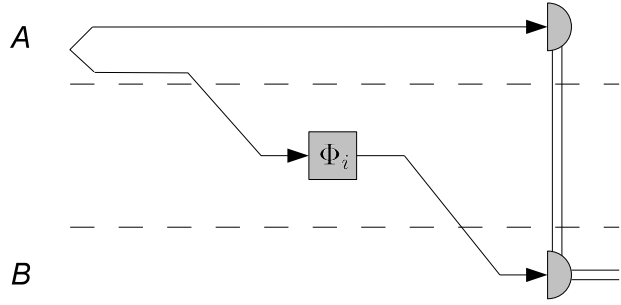} % requires the graphicx package
   \caption{Operational scenario for the LOCC norm. Alice create an entangled probe/ancilla state, keeps the ancilla, and sends the probe down the channel. Alice and Bob can then proceed to an LOCC measurement of the output probe/ancilla state.}
   \label{fig:LOCC}
\end{figure}

Since we focus on families of measurements defined with respect to locality, and the input states are always optimized, three operational minimum-error channel discrimination scenarios corresponding to the three norms $\|\cdot\|_{\textrm{NE}}$, $\|\cdot\|_\diamondsuit$ and $\|\cdot\|_\textrm{LOCC}$ can be depicted as in Figure \ref{fig:trace}, Figure \ref{fig:diamond}, and Figure \ref{fig:LOCC}, respectively. The channels go from Alice to Bob, as in a communication scenario, but the task is not that of transmitting a classical or a quantum message from Alice to Bob having at disposal many uses of the channel. Rather, they want to discriminate between two possible channels with minimum probability of error, having at their disposal only one use of the channel.

In the NE norm case, Alice feeds the channels with a probe that is not correlated with any other subsystems held by either herself or Bob, and Bob measures the output probe, guessing which channel was applied.
 In the diamond-norm case, at the beginning Alice holds the probe and Bob the ancilla, which are in whatever needed pre-distributed entangled state; Alice sends the probe down the channel, and Bob can jointly measure the output probe and the ancilla.\footnote{%
 In this picture, this is a local measurement, as both the output probe and the ancilla are in Bob's lab, but it is a global probe/ancilla measurement.}
 Finally, in the LOCC-norm case, we can imagine that Alice locally creates an entangled probe/ancilla state, feeding the channel with the probe and keeping the ancilla. The measurement is then performed by LOCC on the output probe held by Bob and the ancilla held by Alice.
We remark that this scenario---where the unknown channel $\Phi_i$ goes from Alice to Bob, and it is not in the hands of either of them---could be thought as corresponding to a practical situation like that of, e.g., an optical fiber underneath the sea. In particular, the restriction to LOCC output measurement would be the result of the non-availability of long-distance pre-established entanglement---or of some other quantum channel to be used to send also the ancilla from Alice to Bob. Even more practically, the restriction to LOCC measurement could well capture the possible difficulty in measuring globally the output probe and ancilla in, e.g., a quantum optics setting.

%--------------------------------------------------------------------------%
\section{Optimal discrimination procedures with LOCC measurements}
\label{sec:useful}
%--------------------------------------------------------------------------%

In this section we provide examples of channels $\Phi_0$ and $\Phi_1$
for which
\begin{equation}
  \label{eq:NE-less-than-LOCC}
  \norm{\Phi_0 - \Phi_1}_{\textrm{NE}} < 
  \norm{\Phi_0 - \Phi_1}_{\textrm{LOCC}} =
  \norm{\Phi_0 - \Phi_1}_{\diamondsuit}.
\end{equation}
Thus, entanglement between the input and ancillary systems may still
be useful for channel discrimination when the measurement is
LOCC---and indeed an LOCC measurement may even be optimal in this
situation.

The first collection of examples we provide achieves a limited gap
between the quantities $\norm{\Phi_0 - \Phi_1}_{\textrm{NE}}$ and 
$\norm{\Phi_0 - \Phi_1}_{\textrm{LOCC}}$, but has the advantage of
being simple to describe.
The second type of example achieves a large gap, but also requires
that the output dimension of the channels be very large.
The third example is specifically for input qubits, and while being similar to the
second class of examples, we include it because of its simplicity: on the one hand, it is possible to provide exact values for the the various norms; on the other hand,
it could be experimentally implemented.

%--------------------------------------------------------------------------%
\subsection{Examples based on flagged Kraus operators}
%--------------------------------------------------------------------------%

Let $\X = \complex^d$ for any desired choice of $d\geq 2$, and
let $\Y = \complex^n\otimes\X$ for $n\geq 1$.
Consider a choice of channels $\Phi_0,\Phi_1\in\trans{\X,\Y}$ defined
as
\[
\Phi_a(\rho) = \sum_{j = 1}^n \ket{j}\!\bra{j} \otimes 
A_{a,j} \rho A_{a,j}^{\ast},
\]
for some selection of operators
$\{A_{a,j}\,:\,a\in\{0,1\},\,1\leq j\leq n\}\subset\lin{\X}$ satisfying
\[
\sum_{j = 1}^n A_{0,j}^{\ast}A_{0,j}
=\sum_{j = 1}^n A_{1,j}^{\ast}A_{1,j} = \I_{\X}.
\]

A sufficient condition for $\Phi_0$ and $\Phi_1$ to be perfectly
distinguishable using an LOCC measurement discrimination procedure
(i.e., $\norm{\Phi_0 - \Phi_1}_{\textrm{LOCC}}=2$) is that
$\ip{A_{0,j}}{A_{1,j}} = 0$ for all $j = 1,\ldots,n$.
This observation follows from the well-known theorem of
\cite{WalgateSHV00} stating that every fixed pair of orthogonal pure
states can be perfectly distinguished by an LOCC measurement.
In particular, if the maximally entangled state
\[
\frac{1}{\sqrt{d}}\sum_{i = 1}^d \ket{i}\ket{i}
\]
between the input and an ancillary system is selected, then the
resulting state produced by $\Phi_a$ has the form
\[
\frac{1}{d}\sum_{j = 1}^n \ket{j}\bra{j} \otimes
\ket{\psi_{a,j}}\bra{\psi_{a,j}}
\]
for vectors $\{\ket{\psi_{a,j}}\}$ satisfying
$\langle \psi_{0,j} | \psi_{1,j}\rangle = \ip{A_{0,j}}{A_{1,k}} = 0$
for every choice of $j = 1,\ldots, n$.
An LOCC measurement that first measures $j$, then implements the
corresponding measurement of \cite{WalgateSHV00} to distinguish
$\ket{\psi_{0,j}}$ and $\ket{\psi_{1,j}}$ succeeds in discriminating
$\Phi_0$ and $\Phi_1$ without error.

On the other hand, it holds that
\begin{equation} \label{eq:flagged-NE-norm}
\norm{\Phi_0 - \Phi_1}_{\textrm{NE}}
= \max_{\rho\in\density{\X}}
\sum_{j = 1}^n \norm{A_{0,j} \rho A_{0,j}^{\ast} -
A_{1,j} \rho A_{1,j}^{\ast}}_1.
\end{equation}
So, to obtain examples of channel pairs for which
\eqref{eq:NE-less-than-LOCC} holds, it suffices to select a collection
of operators $\{A_{a,j}\}$ so that
\begin{mylist}{\parindent}
\item[1.] $\ip{A_{0,j}}{A_{1,j}} = 0$ for each $j = 1,\ldots,n$, and
\item[2.] the expression in \eqref{eq:flagged-NE-norm} is smaller
  than~2.
\end{mylist}
This may be accomplished, for instance, by setting $n = d^2 - 1$,
$A_{0,j} = \I/\sqrt{n}$, and $A_{1,j} = W_j/\sqrt{n}$, for 
$j = 1,\ldots, n$ and where $W_1,\ldots,W_n$ is any orthonormal
collection of traceless unitary operators, such as the non-identity
\emph{discrete Weyl} (or \emph{generalized Pauli}) \emph{operators}.
We then have
\[
\norm{\Phi_0 - \Phi_1}_{\textrm{NE}}
= \max_{\ket{\psi}}
\frac{2}{d^2-1}\sum_{j = 1}^{d^2 - 1}
\sqrt{1 - \abs{\bra{\psi}W_j\ket{\psi}}^2}
\leq 2 \sqrt{\frac{d}{d+1}}
\]
where the inequality is due to the concavity of the square root (with equality if and only if Zauner's conjecture
\cite{Zauner99,Appleby05} holds for dimension $d$), and we used the fact that $\rho+\sum_{j=1}^{d^2-1} W_j\rho W_j^* = d \tr(\rho)\I_\cX$

%--------------------------------------------------------------------------%
\subsection{Examples based on random binary-valued measurements}
%--------------------------------------------------------------------------%
\label{sec:example_binary}

\begin{figure}[htbp]
   \centering
   \includegraphics[width=0.3\textwidth]{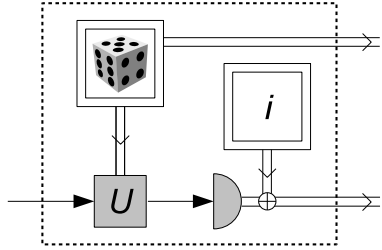} % requires the graphicx package
   \caption{Structure of some exemplary pairs of channels
 %with quantum input and classical output 
 for which the LOCC norm is equal to the diamond norm, and strictly greater than the NE norm. Single lines indicate quantum systems or messages; double lines classical ones. The channel chooses at random a unitary among a discrete set. An incomplete projective measurement in the computational basis with two possible outcomes is then performed. The result of the measurement is further flipped or not, depending on the channel (on the value of $i$ in the figure). The output of the channel is classical, that is diagonal in a fixed computational basis: a register containing the index of the chosen unitary, and a bit containing the (possibly flipped) result of the measurement.  A full description of the action of the channels can be found in the main text.}
   \label{fig:box_with_measurement}
\end{figure}

Let $d$ be an even positive integer, let $\X = \complex^d$ and
$\Y = \complex^2$, and define channels
$\Psi_0,\Psi_1\in\trans{\X,\Y}$ as
\begin{align*}
  \Psi_0(\rho) & = \tr(\Pi_0 \rho) \ket{0}\bra{0} +  \tr(\Pi_1 \rho)
  \ket{1}\bra{1}\\
  \Psi_1(\rho) & = \tr(\Pi_1 \rho) \ket{0}\bra{0} +  \tr(\Pi_0 \rho)
  \ket{1}\bra{1},
\end{align*}
where
\[
\Pi_0 = \sum_{j = 1}^{d/2}\ket{j}\bra{j}
\quad\quad\text{and}\quad\quad
\Pi_1 = \sum_{j = d/2 + 1}^{d}\ket{j}\bra{j}.
\]
Now suppose that $U_1,\ldots,U_N\in\unitary{\X}$ is a collection of
unitary operators, and define channels
$\Phi_0,\Phi_1\in\trans{\X,\complex^N\otimes\Y}$ as follows:
\begin{align*}
  \Phi_0(\rho) & = \frac{1}{N}\sum_{j = 1}^N \ket{j}\bra{j} \otimes
  \Psi_0\!\left(U_j \rho U_j^{\ast}\right)\\
  \Phi_1(\rho) & = \frac{1}{N}\sum_{j = 1}^N \ket{j}\bra{j} \otimes
  \Psi_1\!\left(U_j \rho U_j^{\ast}\right).
\end{align*}
A representation of the action of the channels is provided in Figure \ref{fig:box_with_measurement}.

For any choice of unitary operators $U_1,\ldots,U_N$, the channels
$\Phi_0$ and $\Phi_1$ can be perfectly distinguished by an LOCC
discrimination procedure as follows.
\begin{mylist}{\parindent}
\item[1.]  
  The given channel $\Phi_a$ is evaluated on half of the canonical
  maximally entangled state
  \[
  \ket{\psi} = \frac{1}{\sqrt{d}}\sum_{i = 1}^d\ket{i}\ket{i}.
  \]
  Let us suppose that the resulting state is stored in the registers
  $\reg{A}_1$, $\reg{A}_2$ and $\reg{B}$, where the vector spaces
  corresponding to these registers are given by $\complex^N$,
  $\Y$, and $\X$, respectively; and where Alice holds $\reg{A}_1$ and
  $\reg{A}_2$, and Bob holds $\reg{B}$.

\item[2.] 
  Alice measures $\reg{A}_1$ with respect to the standard basis and
  transmits the result to Bob.
  Upon receiving a value $j\in\{1,\ldots,N\}$, Bob applies the
  operation $\overline{U_j}$ to $\reg{B}$, where the complex
  conjugation is taken with respect to the standard basis.
  The state of the pair $(\reg{A}_2,\reg{B})$ at this point is given
  by $(\Psi_a\otimes\I)(\ket{\psi}\bra{\psi})$.

\item[3.]
  Bob applies $\Psi_0$ to $\reg{B}$ and measures the resulting qubit
  with respect to the standard basis.
  Alice measures her qubit $\reg{A}_2$ with respect to the standard
  basis as well, and Alice and Bob compare their measurements.
  If they agree then $a = 0$ and otherwise $a = 1$.
\end{mylist}
Thus, it holds that $\norm{\Phi_0 - \Phi_1}_{\textrm{LOCC}} = 2$.

Now we will argue that for some choice of unitary operators
$U_1,\ldots,U_N$ it holds that
\[
\norm{\Phi_0 - \Phi_1}_\textrm{NE} = \sqrt{\frac{2}{\pi d}} +
O\left(\frac{1}{d}\right).
\]
This is done by taking $N$ to be large and selecting $U_1,\ldots,U_N$
independently according to the normalized Haar measure on
$\unitary{\X}$.
It holds that
\beq
\label{eq:NE_for_random_binary}
\norm{\Phi_0 - \Phi_1}_{\textrm{NE}}
= 
\max_{\ket{\psi}}
\frac{2}{N}
\sum_{k = 1}^N
\abs{
\sum_{j = 1}^{d/2} \abs{\bra{\psi}U_k\ket{j}}^2 -
\sum_{j = d/2+1}^d \abs{\bra{\psi}U_k\ket{j}}^2},
\eeq
which, in the limit of large $N$, approaches
\[
2\,
\max_{\ket{\psi}}
\int
\abs{
\sum_{j = 1}^{d/2} \abs{\bra{\psi}U_j\ket{j}}^2 -
\sum_{j = d/2+1}^d \abs{\bra{\psi}U_j\ket{j}}^2}
\op{d}\!\mu(U)
\]
for $\mu$ denoting the Haar measure on $\unitary{\X}$, normalized so
that $\mu(\unitary{\X}) = 1$.
A formula for this type of integral is given in Appendix B
of~\cite{MatthewsWW09}, which in this case evaluates to 
\[
\frac{1}{d}\sum_{j=1}^{d/2} 2^{-2j}\binom{2j}{j} 
= \sqrt{\frac{2}{\pi d}} + O\left(\frac{1}{d}\right)
\]
as claimed.

%--------------------------------------------------------------------------%
\subsection{Simple pairs of channels with qubit input}
%--------------------------------------------------------------------------%

Consider a special case of the exemplary class presented in Section \ref{sec:example_binary}, with qubit input dimension $d=2$. In particular, take the number of random unitaries $N=2$, with $U_1=\I$ and $U_2=(\sigma_x+\sigma_z)/\sqrt{2}$. Then Eq. \eqref{eq:NE_for_random_binary} becomes
\[
\label{eq:lowdim}
\norm{\Phi_0 - \Phi_1}_{\textrm{NE}}
= 
\max_{\ket{\psi}}
\sum_{k = x,z}
\abs{
\bra{\psi}\sigma_k\ket{\psi}
}
=\sqrt{2}
\]
where the maximum is realized by a pure state in the $xz$ plane of the Bloch sphere.

This qubit-input example can also be computed exactly for $N=3$, with $U_1=\I,U_2=(\sigma_x+\sigma_z)/\sqrt{2}$ and $U_3=(\sigma_y+\sigma_z)/\sqrt{2}$. Then one has $\norm{\Phi_0 - \Phi_1}_{\textrm{NE}}=2/\sqrt{3}$, with a slightly bigger gap between the LOCC and diamond norms, still both equal to 2, and the NE norm.

%We remark that, although the input of the maps is two-dimensional, the output dimension is higher-dimensional---though classical. Indeed, according to Theorem \ref{thm:nogo_one_qubit} it would have been impossible to find a gap if the output had also been two-dimensional.

%--------------------------------------------------------------------------%
\section{Sub-optimality of LOCC measurements} \label{sec:useless}
%--------------------------------------------------------------------------%

In this section we show that there are cases where restricting
measurements to be LOCC in channel discrimination procedures
eliminates the benefits of entanglement altogether.
That is, there exist channels $\Phi_0$ and $\Phi_1$ for which
\begin{equation} \label{eq:useless}
\norm{\Phi_0 - \Phi_1}_{\textrm{NE}}
= \norm{\Phi_0 - \Phi_1}_{\textrm{LOCC}}
< \norm{\Phi_0 - \Phi_1}_{\diamondsuit}
\end{equation}
Three sets of examples are discussed.

%--------------------------------------------------------------------------%
\subsection{Channels with one-qubit output}
%--------------------------------------------------------------------------%

For any pair of channels whose outputs correspond to a single qubit,
discrimination strategies using LOCC measurements can be no better
than strategies that use no entanglement.
This follows from the following theorem.

\begin{theorem}
  For any choice of channels $\Phi_0,\Phi_1\in\trans{\X,\Y}$ it holds
  that
  \[
  \norm{\Phi_0 - \Phi_1}_{\textup{SEP}} 
  \leq \frac{\dim(\Y)}{2} \norm{\Phi_0 - \Phi_1}_{\textup{NE}}.
  \]
\end{theorem}

\begin{proof}
  Let $\Phi = \Phi_0 - \Phi_1$, and let $\Z = \complex^n$ for an
  arbitrary choice of $n\geq 1$.
  Also let $\rho\in\density{\X\otimes\Z}$ be a density operator and let
  $\{Q_1\otimes R_1,\ldots,Q_N\otimes R_N\}$ represent an arbitrary
  product measurement over $\Y\otimes\Z$, so that
  $Q_1,\ldots,Q_N\in\pos{\Y}$ and $R_1,\dots,R_N\in\pos{\Z}$ satisfy
  \[
  \sum_{j = 1}^N Q_j \otimes R_j = \I_{\Y\otimes \Z}.
  \]
  There is no loss of generality in assuming that each operator $Q_j$
  is a density operator, because each $R_j$ can be re-scaled
  appropriately.

  Now, given that we assume that $Q_1\ldots,Q_N$ are density
  operators, it holds that
  \[
  \sum_{j = 1}^N R_j = \dim(\Y) \I_{\Z}.
  \]
  Our goal will be to establish an upper bound on the quantity
  \begin{equation} \label{eq:product-measurement-sum}
    \sum_{j = 1}^N 
    \abs{\ip{Q_j\otimes R_j}{(\Phi\otimes\I_{\lin{\Z}})(\rho)}}.
  \end{equation}
  To this end let us write 
  $X_j = \tr_{\Z} [(\I_{\Y} \otimes R_j)\rho]$, so that 
  \[
  \abs{\ip{Q_j\otimes R_j}{(\Phi\otimes\I_{\lin{\Z}})(\rho)}}
  = 
  \abs{\ip{Q_j}{\Phi(X_j)}} \leq
  \frac{1}{2}\norm{\Phi}_{\textup{NE}}\tr(X_j)
  \]
  for each $j = 1,\ldots,N$, where the inequality follows from the
  observations that each $\Phi(X_j)$ is traceless and that
  $Q_j \leq \I_{\Y}$.
  It follows that
  \begin{equation} \label{eq:product-measurement-bound}
    \sum_{j = 1}^N 
    \abs{\ip{Q_j\otimes R_j}{(\Phi\otimes\I_{\lin{\Z}})(\rho)}}
    \leq \frac{1}{2} \norm{\Phi}_{\textup{NE}}\sum_{j = 1}^N \tr(X_j)
    = \frac{\dim(\Y)}{2} \norm{\Phi}_{\textup{NE}}.
  \end{equation}
  
  Finally, we note that the quantity $\norm{\Phi}_{\textup{SEP}}$ is
  given by the supremum of the expression
  \eqref{eq:product-measurement-sum} over all choices of $n$ and all
  product measurements $\{Q_1\otimes R_1,\,\ldots,\,Q_N\otimes R_N\}$,
  which completes the proof.
\end{proof}

\begin{corollary}
  Suppose $\Phi_0,\Phi_1\in\trans{\X,\Y}$ are channels for 
  $\dim(\Y) = 2$. 
  Then $\norm{\Phi}_{\textup{NE}}= \norm{\Phi}_{\textup{SEP}}$.
\end{corollary}

As there exist channels $\Phi_0$ and $\Phi_1$ having single-qubit
outputs for which 
$\norm{\Phi_0 - \Phi_1}_{\textup{NE}} <
\norm{\Phi_0 - \Phi_1}_{\diamondsuit}$, we have that
\eqref{eq:useless} holds for these channels.

%--------------------------------------------------------------------------%
\subsection{Werner-Holevo channels}
%--------------------------------------------------------------------------%

The Werner-Holevo channels $\Phi_0$ and $\Phi_1$ are defined for any
dimension $d\geq 2$ as
\[
\Phi_0(X) = \frac{1}{d+1}\left(\tr(X)\I + X^{\t}\right)
\quad\text{and}\quad
\Phi_1(X) = \frac{1}{d-1}\left(\tr(X)\I - X^{\t}\right),
\]
where transposition is taken with respect to the standard basis
\cite{WernerH02}.
It holds that
\[
\norm{\Phi_0 - \Phi_1}_{\textrm{NE}} = \frac{4}{d+1},
\]
whereas $\norm{\Phi_0 - \Phi_1}_{\diamondsuit} = 2$; the channels are
almost indistinguishable for large $d$ without the use of
entanglement, but are perfectly distinguishable with entanglement.

We will now prove that for this choice of channels, LOCC measurements
render entanglement useless for their discrimination.
Indeed, even PPT measurements have this property:
\[
\norm{\Phi_0 - \Phi_1}_\textrm{LOCC} 
= \norm{\Phi_0 - \Phi_1}_\textrm{PPT} 
= \frac{4}{d+1}.
\]

To prove this, we first take $\Z$ to be a Hilbert space having
dimension at least as large as $\X$.
Then any unit vector $\ket{\psi}\in\X\otimes\Z$ may be written
\[
\ket{\psi} = (\I_{\X}\otimes A)\sum_{j = 1}^d \ket{j}\ket{j}
\]
for some choice of a linear mapping $A:\X\rightarrow\Z$ satisfying
$\norm{A}_2 \defeq \sqrt{\ip{A}{A}} = 1$.
Thus, for any mapping $\Phi\in\trans{\X,\Y}$ it holds that
\[
(\Phi\otimes\I_{\lin{\Z}})(\ket{\psi}\bra{\psi})
=(\I_{\Y}\otimes A)J(\Phi)(\I_{\Y}\otimes A)^{\ast},
\]
where
\[
J(\Phi) = \sum_{1\leq j,k\leq d} \Phi(\ket{j}\bra{k})\otimes
\ket{j}\bra{k}
\]
is the \emph{Choi-Jamio{\l}kowski representation} of $\Phi$.
For the mapping $\Phi = \Phi_0 - \Phi_1$, it holds that
\[
J(\Phi_0-\Phi_1) = \frac{2}{d+1}S- \frac{2}{d-1}R,
\]
for $R$ and $S$ denoting the projections onto the anti-symmetric and
symmetric subspaces of $\Y\otimes\X = \complex^d\otimes\complex^d$,
respectively, so that
\[
\norm{\Phi_0 - \Phi_1}_{\textrm{PPT}}
= 2 \ip{(\I_{\Y}\otimes A^{\ast})P(\I_{\Y}\otimes A)}
   {\frac{2}{d-1}R - \frac{2}{d+1}S}
\]
for some choice of $A$ with $\norm{A}_2 = 1$ and some PPT operator
$P\leq \I_{\Y}\otimes\I_{\X}$.
It holds that $\ip{Q}{R - S} \leq 0$ for every PPT operator $Q$,
and $\tr_{\Y}(R) = \frac{d-1}{2}\I_{\X}$, so
that
\[
\norm{\Phi_0 - \Phi_1}_{\textrm{PPT}}
\leq \frac{8}{d^2 - 1} \ip{(\I_{\Y}\otimes A^{\ast})P(\I_{\Y}\otimes
  A)}{R}
\leq \frac{8}{d^2 - 1}\ip{\I_{\Y}\otimes A^{\ast}A}{R} = \frac{4}{d+1}
\ip{A^{\ast} A}{\I_{\X}}.
\]
Given that $A^{\ast} A$ is a density operator, it follows
that
$\norm{\Phi_0 - \Phi_1}_{\textup{PPT}} \leq 4/(d+1)$ as claimed.

%--------------------------------------------------------------------------%
\subsection{The channels of \cite{PianiW09}} 
%--------------------------------------------------------------------------%

Finally, we prove that channels $\Psi_0$ and $\Psi_1$ constructed 
in Theorem 1 of~\cite{PianiW09} satisfy
\[
\norm{\Psi_0-\Psi_1}_\textup{NE} = \norm{\Psi_0-\Psi_1}_\textup{SEP}
< \norm{\Psi_0-\Psi_1}_\diamondsuit.
\]
Channels constructed in this way demonstrate that every entangled
state is useful for channel discrimination, provided that the
final measurement is unrestricted---but by the above relationship we
see that the advantage is lost when the final measurements are
restricted to be separable.

Let us first recall that channels $\Psi_0$ and $\Psi_1$ constructed in
\cite{PianiW09} satisfy $\Psi_0 - \Psi_1 = \alpha \Phi$ for some $\alpha > 0$
and $\Phi\in\trans{\X,\Y\oplus\complex}$ being a mapping defined as
\[
\Phi(X) 
= \begin{pmatrix}
  \Xi(X) & 0\\ 0 & -\tr(X)
\end{pmatrix}
\]
for $\Xi\in\trans{\X,\Y}$ being positive (but not completely positive)
and trace-preserving.
(The mapping $\Phi$ is therefore \emph{trace-annihilating}, which is a
necessary and sufficient condition for the expression
$\Psi_0 - \Psi_1 = \alpha \Phi$ to hold for channels $\Psi_0$ and
$\Psi_1$ and a scalar $\alpha > 0$.)

Now, it is not difficult to see that
$\norm{\Phi}_\textup{SEP} = \norm{\Xi}_\textup{SEP} + 1$.
Specifically, the inequality
$\norm{\Phi}_\textup{SEP} \leq \norm{\Xi}_\textup{SEP} + 1$ follows
from the triangle equality, while the reverse inequality
$\norm{\Phi}_\textup{SEP} \geq \norm{\Xi}_\textup{SEP} + 1$ is
obtained by considering the separable measurement $\{Q_0,Q_1\}\subset\sep{(\cY\oplus\complex):\cX}$
defined as
\[
Q_0 = 
\left(\!\!\!\begin{array}{cl}
P_0 & 0\\0 & 0_\cX
\end{array}
\!\!\!\right)
\quad\quad
\text{and}
%Q_0 = 
%\begin{pmatrix}
%P_0 & 0\\0 & 0_\cX
%\end{pmatrix}
%\quad\quad
%\text{and}
\quad\quad
Q_1 = 
\left(\!\!\!
\begin{array}{cl}
P_1 & 0\\0 & \I_\cX
\end{array}
\!\!\!\right)
%Q_1 = 
%\begin{pmatrix}
%P_1 & 0\\0 & \I_\cX
%\end{pmatrix}
\]
for $\{P_0,P_1\}\subset\sep{\cY:\cX}$ being an optimal separable measurement in the
definition of $\norm{\Xi}_\textup{SEP}$.

Now, for a generic state $\rho\in \density{\X\otimes\Z}$, one finds 
\begin{align*}
  \norm{(\Xi\otimes \I_{\lin{\Z}})(\rho)}_{\textup{SEP}}
  & = \ip{P_0 - P_1}{(\Xi\otimes \I_{\lin{\Z}})(\rho)}\\
  & = \ip{
    (\Xi^{\ast}\otimes \I_{\lin{\Z}})(P_0)
    - (\Xi^{\ast}\otimes \I_{\lin{\Z}})(P_1)}{\rho}
\end{align*}
for some choice of a separable measurement $\{P_0,P_1\}$.
Given that $\Xi$ is positive and trace-preserving we have that
the adjoint mapping $\Xi^{\ast}$ is positive and unital, and given
that $\{P_0,P_1\}$ is separable it therefore follows that
$\{(\Xi^{\ast}\otimes \I_{\lin{\Z}})(P_0),(\Xi^{\ast}\otimes
\I_{\lin{\Z}})(P_1)\}$ is also a (separable) measurement.
The value $\snorm{(\Xi\otimes \I_{\lin{\Z}})(\rho)}_{\textup{SEP}}$ is
therefore equal to 1 (where the equality follows by choosing the
trivial measurement $P_0 = \I$ and $P_1 = 0$, for instance).

Consequently, we have that
$\norm{\Psi_0 - \Psi_1}_{\textup{SEP}} = 2\alpha$.
As is proved in \cite{PianiW09}, it holds that
\[
2\alpha = \norm{\Psi_0 - \Psi_1}_{\textup{NE}} <
\norm{\Psi_0 - \Psi_1}_{\diamondsuit},
\]
and so we have
$\norm{\Psi_0 - \Psi_1}_{\textup{NE}}
= \norm{\Psi_0 - \Psi_1}_{\textup{SEP}}
< \norm{\Psi_0 - \Psi_1}_{\diamondsuit}$
as claimed.

%--------------------------------------------------------------------------%
\section{Conclusions} \label{sec:conclusion}
%--------------------------------------------------------------------------%

We have introduced a new norm $\norm{\cdot}_{\textup{LOCC}}$
on Hermiticity-preserving maps, representing a ``weakened'' version
of the diamond norm.
This norm is motivated by an interest in the usefulness of
entanglement for channel discrimination, and in particular whether
entanglement in the input/ancillary systems remains useful when the
final measurement is implementable by local operations and classical
communication alone.
We provided examples of pairs of channels where input entanglement is
useful only if general measurements are available, while there are
other cases where the advantage of using input entanglement is fully
maintained in spite of the LOCC restriction on measurements.

%We have proved that the particular construction in \cite{PianiW09} will
%not extend to LOCC-restricted measurements, so a different approach
%will be needed to answer this question.

In the context of bipartite state discrimination, one may study the
gap in the probability of error associated to discrimination protocols
that are or are not restricted (e.g., to be LOCC)~\cite{MatthewsWW09}. 
This gap has been shown to depend on the total dimension of the underlying
systems. 
In the present paper we showed two results that regard dimensions. 
One is that a restriction on the output measurement makes input
entanglement useless if the output dimension is too small (one qubit),
and the second is a dimensional lower bound on the gap that can be
achieved between the diamond norm and the LOCC norm, where the
relevant dimension is in this case the input one.
One may consider in more detail what is the role of the input and
output dimensions for the usefulness of entanglement when measurements
are restricted.

In~\cite{PianiW09} it was proved that, for any entangled state, there
exists an instance of a channel discrimination problem for which the
entangled state allows for a correct discrimination with strictly
higher probability than every possible separable state. 
The result was obtained by resorting to the use of the fundamental
characterization of entanglement in terms of linear
maps~\cite{HorodeckiHH96}. 
The actual construction used in~\cite{PianiW09} completely fails when the
output measurement is of the separable type. 
Indeed, in our view the main open problem concerning channel discrimination
with LOCC measurements is whether the results of \cite{PianiW09} can be
extended to LOCC-restricted measurements.
While we know that there are entangled input states useful
for channel discrimination even with constrained output measurements,
the possibility is left open that some entangled states become
useless in this setting.  If this is the case, it would be interesting to understand what kind
of features make some states remain useful and some not.

An issue that has been the focus of a large part of the literature on
the role of entanglement in channel discrimination is the question of
what channels are better distinguished by the used of
entanglement. Similarly, one can ask what channels that are better
discriminated by the use of an entangled ancilla are still (or are not
any more) better discriminated when we impose constraints on the
measurements. The results presented in this work show that both
situations occur. 
Is it possible to find large classes of channels for which input
entanglement stays or fails to stay a useful resource?

%--------------------------------------------------------------------------%
\section*{Acknowledgement}
%--------------------------------------------------------------------------%

The authors would like to thank F.~G.~S.~L.~Brand\~ao and T. Ito for
discussions. 
WM acknowledges the support of the NSERC and QuantumWorks,
MP acknowledges support from NSERC, QuantumWorks, and Ontario Centres of
Excellence, and
JW acknowledges support from NSERC, CIFAR, and QuantumWorks.

%\bibliographystyle{alpha}
%\bibliography{Broken-Diamond}

\newcommand{\etalchar}[1]{$^{#1}$}

\end{document}